# WebCVTree4: A Newly Designed Phylogenetic and Taxonomic Study Platform for Prokaryotes Using Composition Vectors and Whole Genomes

Guanghong Zuo[1,a,*]

[1]*Wenzhou institute, University of Chinese Academy of Sciences, Wenzhou, Zhejiang 325001, China*

*\* Corresponding author(s).*
E-mail: ghzuo@ucas.ac.cn (Zuo GH).

**Running title:** *Zuo GH / WebCVTree4 Servers*

[a] ORCID: 0000-0002-7822-5969.

Total figures: 3

Total supplementary files: 1

**Abstract**

CVTree is an alignment-free methodology for inferring species phylogeny and taxonomy. This method allows for the efficient and accurate resolution of evolutionary relationships among large numbers of species based on whole-genome sequence data. Since 2004, we have been continuously providing CVTree web services. Recently, the server has undergone a significant upgrade, culminating in the release of the WebCVTree4 platform. This upgrade encompasses a comprehensive update of the inbuilt genomic database. Concurrently, the core algorithm has been optimized to support online phylogenetic reconstruction for tens of thousands of species, thereby facilitating the construction of genome-based trees of life. Moreover, we have developed a novel algorithm for comparing phylogenetic trees with established taxonomic systems. This algorithm allows for rapid tree rooting, taxonomic annotation, and topology comparison. Through an interactive web-based visualization tool, users can dynamically adjust tree layouts and export high-quality phylogenetic tree figures. This functionality provides robust support for comparative analysis between CVTree-generated phylogeny and taxonomy. As genome sequencing costs continue to decline, research into microbial evolution and the revision of taxonomic frameworks will increasingly rely on whole-genome data. WebCVTree4 will serve as an efficient web-based platform to support studies in microbial phylogenetics and taxonomy, accessible at https://cvtree.online/.

**Introduction**

The rapid advancement of high-throughput sequencing technologies and the significant reduction in sequencing costs have led to an unprecedented surge in genomic sequence data worldwide [1–3]. These massive genomic datasets provide unparalleled opportunities for deciphering the origin of life, evolutionary history, and biodiversity, while simultaneously posing substantial challenges to traditional bioinformatic analysis methods [4,5]. Among various bioinformatic analyses, phylogenetic tree reconstruction serves as a fundamental method for elucidating evolutionary relationships among species or genes and remains an indispensable core analytical tool [6,7] in fields such as evolutionary biology [8,9], microbiology [10,11], metagenomics [12], and taxonomy [13,14].

Traditional phylogenetic tree construction methods (e.g., Maximum Parsimony, Maximum Likelihood, and Bayesian Inference) primarily rely on multiple sequence alignment (MSA) [15]. These methods infer evolutionary relationships by aligning homologous sites in nucleotide or protein sequences according to substitution models, and they have achieved remarkable success in resolving many deep evolutionary questions. However, as the number of genomes grows exponentially, their limitations have become increasingly evident: Firstly, performing accurate MSA on distantly related sequences is inherently a computationally complex and highly uncertain task [16]. Secondly, these methods are computationally demanding and time-consuming, making it difficult to meet the requirements for large-scale analyses involving thousands to tens of thousands of genomes [17].

To overcome these bottlenecks, alignment-free phylogenetic reconstruction methods have emerged[18–23]. These approaches regard whole-genome sequences as informational entities, and infer evolutionary relationships by extracting global statistical features from the sequences, thus avoiding the cumbersome and error-prone MSA step. The CVTree method, proposed by Academician Bailin Hao's team, is a mature and representative alignment-free approach for studying species phylogenetic relationships based on whole-genome data [24]. It transforms the study of gene sequences by constructing phylogenetic trees to estimate species relationships and elucidate evolutionary paths. These vectors are generated by calculating k-mer frequencies and eliminating background noise resulting from genetic code degeneracy (predicted using a Markov model). The evolutionary distances between different genomes are quantified by calculating the cosine of the angle between their corresponding composition vectors, and phylogenetic trees are subsequently constructed based on the resulting distance matrix [25]. Since its introduction, the CVTree method has demonstrated it has robust utility across multiple domains, particularly in taxonomic identification of prokaryotes [26–30], tracing the evolutionary history of viruses [31], and resolving deep evolutionary relationships in eukaryotes [32–35]. It provides evolutionary perspectives complementary to classical methods and sometimes offers higher resolution [36,37]. It should be noted that the CVTree method was originally designed for constructing species trees based on whole-genome data. However, our recent studies demonstrate that for 16S rRNA, the phylogenetic trees generated by CVTree achieve performance comparable to traditional methods while exhibiting much faster computational efficiency [38].

To make this powerful tool more accessible to the broader research community, the CVTree project released a publicly accessible web server early on, which has since evolved through three versions [39–41]. The WebCVTree3 server, via its user-friendly web interface, shields users from complex command-line operations and computational details, significantly lowering the barrier to entry and promoting the adoption of the method [41]. However, as scientific research enters a new era characterized by big data and open science, the original web server, after over a decade of service, has gradually revealed limitations such as computational bottlenecks, data obsolescence, and insufficient functionality and interactivity. Concurrently, advancements in web development technologies, cloud computing resources, and high-performance computing libraries provide the technical feasibility to address these issues. Constructing modern, high-performance, user-friendly online bioinformatics platforms has become crucial infrastructure for advancing scientific discovery.

In this article, we introduce the newly upgraded WebCVTree4 web server. Deployed on Aliyun with a dedicated domain (https://cvtree.online/v4/), WebCVTree4, compared with previous versions, features the following key improvements: it is constructed based on the RefSeq database [42], integrating over 200,000 genomes; the newly enhanced algorithms empower it to conduct online phylogenetic reconstruction for tens of thousands of species, thus facilitating the construction of a genome-based tree of life; the introduction of the CLTree algorithm enables rapid phylogenetic rooting, taxonomic annotation, and systematic comparison of a phylogenetic tree against the taxonomic system [43]. Ultimately, these comparison results are displayed through an interactive web interface, which provides tree manipulation and beautification functions, and allows for the export of publication-ready tree figures.

**Algorithm and Implementation**

Throughout the evolutionary development process of WebCVTree, we set three key technical benchmarks: whole-genome analysis, alignment-free methodology, and independent validation. Corresponding solutions were proposed, including whole-genome data integration, the alignment-free CVTree algorithm, and the CLTree algorithm for comparing phylogenetic trees with taxonomy. These solutions were incorporated into WebCVTree4, creating a comprehensive and integrated system. Its workflow is shown in Figure 1 and consists of the following three main modules.

*Built-In Data*

In WebCVTree4, both user-uploaded and system-built-in whole-genome data are used as input for the CVTree program. The server's built-in whole-genome dataset primarily originates from the NCBI RefSeq database [42], while also incorporating partial data from JGI IMG [44] and PATRIC [45,46]. The latest version of WebCVTree4's built-in database currently contains 224,977 genomic entries, including 1154 archaeal genomes, 223,815 bacterial genomes (encompassing 81 Tiny genomes), and 8 eukaryotic genomes (which may serve as outgroup references). The term "Tiny genomes" here refers to microbial genomes whose sequences are very short, predominantly representing highly degenerated genomes of bacterial endosymbionts. Degradation has resulted in the loss of many genes, and they tend to the root and occasionally violate the trifurcation of the three main domains of life in the phylogenetic tree. Therefore, it is recommended to exclude these genomes, except when the study is specifically focused on these endosymbiotic bacteria. And it should be noted that the newly integrated CLTree algorithm enables optimal matching between unrooted trees and taxonomic systems based on taxonomic information and branch length data. Consequently, the outgroup designation here serves solely as an identifier and does not participate in the rooting process of unrooted trees.

In addition to the built-in genomic data, WebCVTree4 integrates taxonomic information, with the core data sourced from the dump files of the NCBI Taxonomy database. For the built-in genomes, taxonomic information was primarily retrieved from the NCBI Taxonomy database [47,48] based on the corresponding NCBI Taxon IDs. This information was subsequently curated and supplemented by cross-referencing the LPSN database [49,50] and incorporating recent taxonomic revisions published in the International Journal of Systematic and Evolutionary Microbiology (IJSEM). For user-uploaded data, the system retrieves provisional taxonomic annotations based on filenames using a localized version of the NCBI taxonomy dump. Moreover, the system offers a user-friendly web-based interactive interface, enabling users to view and modify the taxonomic information for each genome.

*CVTree Approach*

The CVTree method, which is widely used in phylogenetic reconstruction, employs distance matrices to infer evolutionary relationships among organisms. This method begins by counting the frequencies of specific k-mers of length k within genomes. Background noise is subsequently predicted using a Markov model, enabling the construction of a high-dimensional composition vector for each species. Specifically, the composition vector representing the whole genome is defined as:

$$V(a_1 a_2 \cdots a_K) = \frac{p(a_1 a_2 \cdots a_K) - \tilde{p}(a_1 a_2 \cdots a_K)}{\tilde{p}(a_1 a_2 \cdots a_K)}$$

$p(a_1 a_2 \cdots a_K) = \frac{N(a_1 a_2 \cdots a_K)}{N_K}$ represents the occurrence probability of the k-mer $a_1 a_2 \cdots a_K$ and $\tilde{p}(a_1 a_2 \cdots a_K) \approx \frac{p(a_1 \cdots a_{K-1}) \times p(a_2 \cdots a_K)}{p(a_2 \cdots a_{K-1})}$ denotes the prior probability estimated from the occurrence probabilities of the (K-1)-mer and (K-2)-mer based on the Markov model. The genetic distance between species is defined by the cosine of the angle between their composition vectors, expressed mathematically as:

$$d_{ij} = \frac{1}{2}\left(1 - \frac{\boldsymbol{V}_i \cdot \boldsymbol{V}_j}{|\boldsymbol{V}_i| \cdot |\boldsymbol{V}_j|}\right)$$

This genetic distance is normalized, ranging from 0 to 1. The phylogenetic tree is ultimately constructed using the neighbor-joining (NJ) method.

The CVTree algorithm can study species evolution based on DNA, RNA, or protein sequences. Previous studies had shown that using protein sequences yields results more consistent with taxonomy[24]. For the selection of K, i.e. the length of k-mer, a reasonable length is in the range $\log_m N < k < \log_m N + 2$, where $m$ is the number of letters in the type of genome and $N$ is the average length of the genome sequences [51,52]. For protein sequences of prokaryote, $m = 20$ and $N \sim 10^7$, Therefore, protein sequences and $k = 5, 6, 7$ are selected by default in WebCVTree4, where $k = 6$ denotes the optimal k-mer length for most of prokaryotic sequences, while this parameter $k = 5\ and\ 7$ can serve as supplementary references.

An updated CVTree program [25] was used in WebCVTree4. It enhances computational efficiency by querying cached intermediate data, directly accessing pre-computed genome composition vectors and genetic distances to avoid redundant calculations. Consequently, WebCVTree4 caches not only the built-in genome data but also the composition vectors of the core genome and the pairwise dissimilarities between species. Intermediate data (including composition vectors and dissimilarities) generated from user-uploaded genomes are cached within the project directory. This cached data can be reused until the project is deleted, or leveraged for creating new projects via the "Copy Project" function (refer the WebCVTree4 user manual for detail).

*CLTree Approach*

A defining feature introduced in WebCVTree3 is the validation of resultant phylogenetic trees through direct comparison with established taxonomic classifications, as opposed to reliance on statistical resampling techniques like bootstrap. This methodological choice stems from the recognition that bootstrap methods, widely employed in phylogenetic inference, primarily evaluate the consistency of the data with the inferred topology rather than assessing the deterministic accuracy of the proposed evolutionary relationships[51]. Furthermore, the CVTree algorithm inherently produces unrooted trees, which depict relative genetic relatedness without specifying ancestral directions, thereby necessitating a subsequent rooting procedure to interpret evolutionary pathways[53]. In earlier iterations of CVTree, rooting was accomplished via the outgroup method. This technique establishes the root by including operational taxonomic units that are phylogenetically proximate yet evolutionarily external to the clade of primary interest. However, the selection of a suitable outgroup presents considerable challenges in specific scenarios, such as during the reconstruction of a comprehensive tree of life that aims to encompass all extant species. Given that biological evolution and taxonomy are intrinsically interconnected [54–56], the integration of taxonomic knowledge with statistical methodologies for both tree evaluation and rooting offers a novel and substantive framework for investigating evolutionary relationships[57].

In WebCVTree4, we introduced the Collapsed Lineage Tree (CLTree) to root and evaluate the phylogenetic tree by using taxonomic information..[43]. The tool first retrieves the taxonomic information for each genome on the phylogenetic tree via a database query system. It then performs recursive computations starting from the root node of the tree, annotating each node with a taxonomic label based on the lowest common taxonomic unit shared by all descendant genomes. Finally, the congruence between the phylogenetic tree and taxonomic units at various ranks is quantified using taxonomic labels, as discussed in the context of phylogenetic tree construction and validation standards. Algorithmic analysis reveals that the core algorithm of CLTree operates in linear time complexity, enabling efficient processing of large-scale datasets. The algorithm has been packaged as standalone software, with detailed methodologies available in the associated publication [43] and user manual.

**Applications of Web Server**

WebCVTree4 is deployed on an Aliyun Cloud server with a dedicated domain. The latest version of the server can be accessed via the URL https://cvtree.online/v4. Its working interface and navigation logic are illustrated in Figure 2. The system comprises nine core functional pages that implement project configuration, result visualization, and input/output operations. This article merely offers a concise overview of the key enhancements to the new server; meanwhile, detailed operational procedures are documented in the WebCVTree4 user manual.

*Project Configuration Settings*

In Figure 2, the pages related to project configuration are highlighted in green. These primarily include the Start Page for project initialization, basic parameter configuration, genome data configuration (supporting selection from built-in databases or local file upload), and taxonomic information configuration pages.

In the project initialization module, the system features a new "Copy Project" function button. This allows users to copy an existing project and initialize it as a new one. This functionality offers significant practical benefits: it enables users to adjust existing project parameters without full reconfiguration, effectively avoids redundant data uploads, and supports the CVTree program can leverage intermediate results from previous project runs to avoid redundant calculations, while also supporting the preservation and comparative analysis of results from multiple runs. It is worth noting that this feature was introduced in a minor update of the previous version.

In the basic parameter configuration and genome data configuration modules, aside from workflow optimizations, the core CVTree module for generating phylogenetic trees remains stable. The crucial server upgrade in this release primarily focuses on the comparison of phylogenetic trees with taxonomic systems. The system's backend has been enhanced to perform synchronized, regular updates of the NCBI Taxonomy database, ensuring the integration of the latest taxonomic information. Additionally, it incorporates a novel CLTree algorithm to optimize data processing and analysis. This allows for the pre-assignment of taxonomic information to user-uploaded genome datasets and enables efficient comparison between inferred phylogeny and the taxonomy. On the frontend, the system introduces a web-based tabular interface. Users can configure taxonomic information for individual genomes via the table and perform bulk configuration through locally edited CSV files, enabling convenient server interaction.

*Display Result*

After completing system configuration and submitting the project, the time required to construct the phylogenetic tree depends on the number of project genomes and cache status (as shown by the red gear icon in Figure 2). Users may record the project ID and close the page at this stage. The project can be subsequently reloaded by entering the project ID via the Start Page. If an email address is provided in the basic settings, the system will automatically send a notification upon computation completion, allowing users to directly access results via the link in the email.

Result display pages, highlighted in cyan in Figure 2, primarily consist of two visualization modules: lineage and tree perspectives. WebCVTree4 maintains the web-based interactive design philosophy for result presentation. Users can dynamically adjust display modes via click interactions, enabling quick identification of critical information. In addition to the previously implemented interactive phylogenetic tree viewer, this update introduces a hierarchical taxonomic information display scheme within the classification perspective module. This scheme utilizes a collapsible interaction model, greatly improving presentation conciseness and readability.

*Output Result*

As shown in Figure 2, the results output page is highlighted in blue. This functional module has been newly added and optimized according to user feedback from the previous version. To improve computational efficiency, the CVTree program stores intermediate results, such as species dissimilarity data, in binary files during the computation process. Since these files are typically large, intermediate results are excluded from the default downloadable output. Addressing researchers' needs to view specific dissimilarity data, WebCVTree4 added a new page for the selective output of partial species dissimilarity data to support subsequent analysis. Users may also opt to run the local version of the CVTree program [25] (download source code from: https://github.com/ghzuo/CVTree) to obtain the complete set of intermediate results.

This update enhances the phylogenetic tree output functionality by incorporating advanced analytical tools and methodologies. When tree structures contain numerous branches, dynamically adjusting branch quantities and redrawing the tree consumes significant local computational resources. In order to balance efficiency, the phylogenetic tree output process has been optimized into a two-step workflow: Users can first adjust the branch structure of the target phylogenetic tree on the TreeView page; subsequently, they can perform tree beautification on the Output Tree page.

Figure 3 shows an example phylogenetic tree output by WebCVTree4, which comprises 5553 genome sequences (235 archaeal, 5310 bacterial, and 8 eukaryotic genomes, belonging to 41 phyla) and hides the branches with taxonomic rank lower than phylum. On the CVTree-constructed phylogeny, 40 phyla formed monophyletic clades. Notably, even the *Pseudomonadota* phylum, comprising 2324 genomes, exhibited monophyly. However, the phylum *Spirochaetota*, including three classes *Leptospiria*, *Brachyspiria*, and *Spirochaetia,* arranged chronologically, exhibited a polyphyletic distribution in the phylogenetic tree (highlighted in green in Figure 3). It should be noted that the controversy surrounding the phylum *Spirochaetota* has a long history [58]. The phylum *Spirochaetota* is defined based on phenotype (morphology and physiology), and the bacteria are classified into this phylum primarily due to their helical morphology. However their physiology and habitats are heterogeneous [59], and they are diverges in genotype (phylogeny of genetic sequences) [60]. After the latest taxonomic revisions in 2024 [61,62], there are five valid classes in the phylum *Spirochaetota*. These classes were initially classified as genus and were elevated to the rank of class in several taxonomic revisions. We believe that it was not the last elevation for these classes because the current revision has not yet led to the monophyletic classification of Spirochaetota. In fact, as early as 2007, we proposed based on CVTree results that the class *Leptospiria* (as family *Leptospiraceae* at that time) should be elevated to a rank of phylum [63]. There were no genome available from the class *Brachyspiria* in previous study. However, according to the phylogenetic tree in Figure 3, class *Brachyspiria* should also be elevated to the rank of phylum.

**Discussion and Conclusion**

The innovation in current sequencing technologies has generated massive amounts of genomic data, thereby providing abundant resources for whole-genome-based data analysis, which simultaneously presents challenges in developing appropriate computational tools. The computational bottlenecks and biases introduced by manual selection have spurred the development of parameter-free and alignment-free methods. The CVTree method was developed to address this requirement. It systematically employs whole-genome information to build species tree to facilitate investigations into phylogenetic relationships and taxonomic system research. This approach demonstrates substantial concordance with taxonomy [36], while concurrently providing strain-level resolution and assisting in the resolution of classification challenges that entail specific criteria [28,41]. Moreover, recent studies demonstrate that even for gene tree construction, the CVTree method achieves performance comparable to traditional sequence alignment approaches, while exhibiting significantly faster execution speeds [38].

Our newly developed WebCVTree4 web server, which operates on parallel high-performance hardware, serves as an efficient and convenient tool for whole-genome-based phylogenetic and taxonomic investigations. It features a user-friendly interface and enables automated, efficient comparison of phylogenetic relationships and taxonomic classifications. Results are presented from dual perspectives, namely taxonomic classification and phylogenetic trees, and it also allows interactive operations within the browser. Moreover, it generates high-quality tree diagrams for the purpose of visualization. This platform substantially enhances the capacity of researchers in prokaryotic biology to employ the CVTree methodology.

We anticipate that as genome sequencing becomes increasingly accessible, research on prokaryotic phylogenetic relationships and taxonomic systems will increasingly rely on whole-genome-based findings. WebCVTree is poised to become a robust web-based platform for the study of prokaryotic evolution and taxonomy, concurrently establishing the CVTree method as a definitive tool for clarifying prokaryotic phylogeny and taxonomic structures.

**Server availability**

The WebCVTree4 Server is freely available at https://cvtree.online.

**CRediT author statement**

**Zuo Guanghong**: Conceptualization, Methodology, Software, Data curation, Visualization, Investigation, Supervision, Validation, Writing.

**Competing Interests**

The authors have declared that no competing interests exist.

**ORCID**

0000-0002-7822-5969 Guanghong Zuo

**Acknowledgments**

Funding: This work was supported by the Wenzhou institute, University of Chinese Academy of Sciences (Grant No. WIUCASQD2021042).

# References

[1] Stephens ZD, Lee SY, Faghri F, Campbell RH, Zhai C, Efron MJ, et al. Big data: Astronomical or genomical? PLOS Biology 2015;13:e1002195.

[2] Loman NJ, Pallen MJ. Twenty years of bacterial genome sequencing. Nat Rev Microbiol 2015;13:787–94.

[3] Shendure J, Balasubramanian S, Church GM, Gilbert W, Rogers J, Schloss JA, et al. DNA sequencing at 40: Past, present and future. Nature 2017;550:345–53.

[4] Gauthier J, Vincent AT, Charette SJ, Derome N. A brief history of bioinformatics. Brief Bioinform 2019;20:1981–96.

[5] Yang A, Troup M, Ho JWK. Scalability and validation of big data bioinformatics software. Comput Struct Biotechnol J 2017;15:379–86.

[6] Felsenstein J. Inferring phylogenies. Sunderland, Mass: Sinauer Associates is an imprint of Oxford University Press; 2004.

[7] Nei M, Kumar S. Molecular evolution and phylogenetics. Cary: Oxford University Press; 2000.

[8] Johnson KP, Dietrich CH, Friedrich F, Beutel RG, Wipfler B, Peters RS, et al. Phylogenomics and the evolution of hemipteroid insects. Proc Natl Acad Sci 2018;115:12775–80.

[9] Zaremba-Niedzwiedzka K, Caceres EF, Saw JH, Bäckström D, Juzokaite L, Vancaester E, et al. Asgard archaea illuminate the origin of eukaryotic cellular complexity. Nature 2017;541:353–8.

[10] Baker BA, McCarthy CGP, López-García P, Leroy RB, Susko E, Roger AJ, et al. Phylogenomic analyses indicate the archaeal superphylum DPANN originated from free-living euryarchaeal-like ancestors. Nat Microbiol 2025:1–12.

[11] Eme L, Tamarit D, Caceres EF, Stairs CW, De Anda V, Schön ME, et al. Inference and reconstruction of the heimdallarchaeial ancestry of eukaryotes. Nature 2023;618:992–9.

[12] Simmonds P, Adams MJ, Benkő M, Breitbart M, Brister JR, Carstens EB, et al. Virus taxonomy in the age of metagenomics. Nat Rev Microbiol 2017;15:161–8.

[13] Eme L, Spang A, Lombard J, Stairs CW, Ettema TJG. Archaea and the origin of eukaryotes. Nat Rev Microbiol 2017;15:711–23.

[14] Fox GE, Woese CR. Classification of methanogenic bacteria by 16S ribosomal RNA characterization. Proc Natl Acad Sci U S A 1977;74:4537–41.

[15] Yang Z. Molecular evolution: A statistical approach. Oxford (GB): Oxford University Press; 2014.

[16] Edgar RC. MUSCLE: multiple sequence alignment with high accuracy and high throughput. Nucleic Acids Res 2004;32:1792–7.

[17] Stamatakis A. RAxML version 8: a tool for phylogenetic analysis and post-analysis of large phylogenies. Bioinformatics 2014;30:1312–3.

[18] Bonham-Carter O, Steele J, Bastola D. Alignment-free genetic sequence comparisons: a review of recent approaches by word analysis. Briefings Bioinf

2014;15:890–905.

[19] Zielezinski A, Vinga S, Almeida J, Karlowski WM. Alignment-free sequence comparison: benefits, applications, and tools. Genome Biol 2017;18:186.

[20] Ren J, Bai X, Lu YY, Tang KJ, Wang Y, Reinert G, et al. Alignment-free sequence analysis and applications. Annu Rev Biomed Data Sci Vol 1 2018;1:93–114.

[21] Zielezinski A, Girgis HZ, Bernard G, Leimeister CA, Tang KJ, Dencker T, et al. Benchmarking of alignment-free sequence comparison methods. Genome Biol 2019;20.

[22] Vinga S. Information theory applications for biological sequence analysis. Briefings Bioinf 2014;15:376–89.

[23] Vinga S, Almeida J. Alignment-free sequence comparison - a review. Bioinformatics 2003;19:513–23.

[24] Qi J, Wang B, Hao B. Whole proteome prokaryote phylogeny without sequence alignment: a K-string composition approach. J Mol Evol 2004;58:1–11.

[25] Zuo G. CVTree: a parallel alignment-free phylogeny and taxonomy tool based on composition vectors of genomes. Genomics Proteomics Bioinformatics 2021;19:662–7.

[26] Zuo G, Xu Z, Hao B. Shigella strains are not clones of escherichia coli but sister species in the genus escherichia. Genomics Proteomics Bioinformatics 2013;11:61–5.

[27] Zuo G, Xu Z, Hao B. Phylogeny and taxonomy of archaea: a comparison of the whole-genome-based CVTree approach with 16S rRNA sequence analysis. Life (Basel) 2015;5:949–68.

[28] Zuo G, Hao B, Staley JT. Geographic divergence of "sulfolobus islandicus" strains assessed by genomic analyses including electronic DNA hybridization confirms they are geovars. Antonie van Leeuwenhoek 2014;105:431–5.

[29] Zhang Q, Wu Y, Wang J, Wu G, Long W, Xue Z, et al. Accelerated dysbiosis of gut microbiota during aggravation of DSS-induced colitis by a butyrate-producing bacterium. Sci Rep 2016;6:27572.

[30] Liu J, Wang H, Yang H, Zhang Y, Wang J, Zhao F, et al. Composition-based classification of short metagenomic sequences elucidates the landscapes of taxonomic and functional enrichment of microorganisms. Nucleic Acids Res 2013;41:e3.

[31] Gao L, Qi J. Whole genome molecular phylogeny of large dsDNA viruses using composition vector method. BMC Evol Biol 2007;7:41.

[32] Wang H, Xu Z, Gao L, Hao B. A fungal phylogeny based on 82 complete genomes using the composition vector method. BMC Evol Biol 2009;9:195.

[33] Kjaerbolling I, Vesth TC, Frisvad JC, Nybo JL, Theobald S, Kuo A, et al. Linking secondary metabolites to gene clusters through genome sequencing of six diverse aspergillus species. Proc Natl Acad Sci U S A 2018;115:E753–61.

[34] Chu KH, Qi J, Yu ZG, Anh V. Origin and phylogeny of chloroplasts revealed by a simple correlation analysis of complete genomes. Mol Biol Evol 2004;21:200–6.

[35] Yuan J, Zhu Q, Liu B. Phylogenetic and biological significance of evolutionary elements from metazoan mitochondrial genomes. PLOS One 2014;9:e84330.

[36] Zuo G, Qi J, Hao B. Polyphyly in 16S rRNA-based LVTree versus monophyly in whole-genome-based CVTree. Genomics Proteomics Bioinformatics 2018;16:310–9.

[37] Zuo G, Hao B. On monospecific genera in prokaryotic taxonomy. Synth Syst Biotechnol 2017;2:226–35.

[38] Lu Y-F, Zhi X-Y, Zuo G-H. CVTree for 16S rRNA: Constructing taxonomy-compatible all-species living tree effectively and efficiently. Chinese Phys B 2025;34:088704.

[39] Qi J, Luo H, Hao B. CVTree: a phylogenetic tree reconstruction tool based on whole genomes. Nucleic Acids Res 2004;32:W45-7.

[40] Xu Z, Hao B. CVTree update: A newly designed phylogenetic study platform using composition vectors and whole genomes. Nucleic Acids Res 2009;37:W174-8.

[41] Zuo G, Hao B. CVTree3 web server for whole-genome-based and alignment-free prokaryotic phylogeny and taxonomy. Genomics Proteomics Bioinformatics 2015;13:321–31.

[42] O'Leary NA, Wright MW, Brister JR, Ciufo S, Haddad D, McVeigh R, et al. Reference sequence (RefSeq) database at NCBI: current status, taxonomic expansion, and functional annotation. Nucleic Acids Res 2016;44:D733–45.

[43] Zuo G. CLTree: a tool for annotating, rooting, and evaluating phylogenetic trees leveraging genomic lineages 2025.

[44] Chen I-MA, Chu K, Palaniappan K, Ratner A, Huang J, Huntemann M, et al. The IMG/M data management and analysis system v.7: Content updates and new features. Nucleic Acids Res 2023;51:D723–32.

[45] Wattam AR, Abraham D, Dalay O, Disz TL, Driscoll T, Gabbard JL, et al. PATRIC, the bacterial bioinformatics database and analysis resource. Nucleic Acids Res 2014;42:D581–91.

[46] Wattam AR, Davis JJ, Assaf R, Boisvert S, Brettin T, Bun C, et al. Improvements to PATRIC, the all-bacterial bioinformatics database and analysis resource center. Nucleic Acids Res 2017;45:D535–42.

[47] Federhen S. The NCBI taxonomy database. Nucleic Acids Res 2012;40:D136–43.

[48] Sakamoto T, Ortega JM. Taxallnomy: an extension of NCBI taxonomy that produces a hierarchically complete taxonomic tree. BMC Bioinf 2021;22:388.

[49] Parte AC, Sardà Carbasse J, Meier-Kolthoff JP, Reimer LC, Göker M. List of prokaryotic names with standing in nomenclature (LPSN) moves to the DSMZ. Int J Syst Evol Microbiol 2020;70:5607–12.

[50] Parte AC. LPSN—list of prokaryotic names with standing in nomenclature. Nucleic Acids Res 2014;42:D613–6.

[51] Zuo G, Xu Z, Yu H, Hao B. Jackknife and bootstrap tests of the composition vector trees. Genomics Proteomics Bioinformatics 2010;8:262–7.

[52] Zuo G, Li Q, Hao B. On K-peptide length in composition vector phylogeny of

prokaryotes. Comput Biol Chem 2014;53 Pt A:166–73.

[53] Tria FDK, Landan G, Picazo DR, Dagan T. Phylogenomic Testing of Root Hypotheses. Genome Biology and Evolution 2023;15:evad096.

[54] Austin B, Colwell RR. Evaluation of some coefficients for use in numerical taxonomy of microorganisms. Int J Syst Bacteriol 1977;27:204–10.

[55] Chun J, Rainey FA. Integrating genomics into the taxonomy and systematics of the bacteria and archaea. Int J Syst Evol Microbiol 2014;64:316–24.

[56] Hugenholtz P, Chuvochina M, Oren A, Parks DH, Soo RM. Prokaryotic taxonomy and nomenclature in the age of big sequence data. ISME J 2021;15:1879–92.

[57] Matsen FA, Gallagher A. Reconciling taxonomy and phylogenetic inference: formalism and algorithms for describing discord and inferring taxonomic roots. Algorithms Mol Biol 2012;7:8.

[58] Canale-Parola E. Physiology and evolution of spirochetes. Bacteriol Rev 1977;41:181–204.

[59] Johnson RC. Introduction to the spirochetes. In: Starr MP, Stolp H, Trüper HG, Balows A, Schlegel HG, editors. The prokaryotes: a handbook on habitats, isolation, and identification of bacteria, Berlin, Heidelberg: Springer; 1981, p. 533–7.

[60] Abt B, Han C, Scheuner C, Lu M, Lapidus A, Nolan M, et al. Complete genome sequence of the termite hindgut bacterium spirochaeta coccoides type strain (SPN1T), reclassification in the genus sphaerochaeta as sphaerochaeta coccoides comb. nov. and emendations of the family spirochaetaceae and the genus sphaerochaeta. Stand Genomic Sci 2012;6:194–209.

[61] Oren A, Göker M. Validation list no. 215. Valid publication of new names and new combinations effectively published outside the IJSEM. Int J Syst Evol Microbiol 2024;74:6173.

[62] Chuvochina M, Mussig AJ, Chaumeil P-A, Skarshewski A, Rinke C, Parks DH, et al. Proposal of names for 329 higher rank taxa defined in the genome taxonomy database under two prokaryotic codes. FEMS Microbiol Lett 2023;370:fnad071.

[63] 高雷，戚继，孙健冬，郝柏林. 原核生物系统发生学与分类学的一致性：组份矢量树与原核生物分类系统的详尽比较. 中国科学 c 辑 2007;37:389–401.

**Figure Legends**

*Figure 1 Workflow of the Backend of WebCVTree4.*

The backend system of WebCVTree4 comprises three core modules: a built-in module for whole-genome data and taxonomic information; a whole-genome-based module for generating the CVTree phylogenetic tree; and a module for comparing the CLTree phylogenetic tree with taxonomic systems. Moreover, the server backend employs a cache mechanism for intermediate results to optimize computational efficiency.

*Figure 2 Flow Diagram of the Frontend of WebCVTree4.*

Each block represents an independent page, with arrow directions indicating navigation paths between pages. All interactive pages constitute three functional modules: the Project Setup module (depicted in green), the Results Visualization module (depict in cyan), and the Data Output module (depicted in blue). The system is non-interactive during jobs execution (denoted by white gears in red cloud icon). Once the job is completed, users can explore the analysis results via the interactive interface. Finally, users may export optimized phylogenetic trees for visualization or download distance matrices for subsequent in-depth analysis.

*Figure 3 Example for Output Tree by WebCVTree4*

There are 5,553 gene sequences in total, including 235 archaeal genomes (depicted in red), 5,310 bacterial genomes (depicted in blue for monophyly phyla and green for non-monophyly phyla), and 8 eukaryotic genomes (depicted in terracotta). All phylum names are annotated in the figure, and the number of genomes analyzed for each phylum in this study is provided in parentheses after the name.

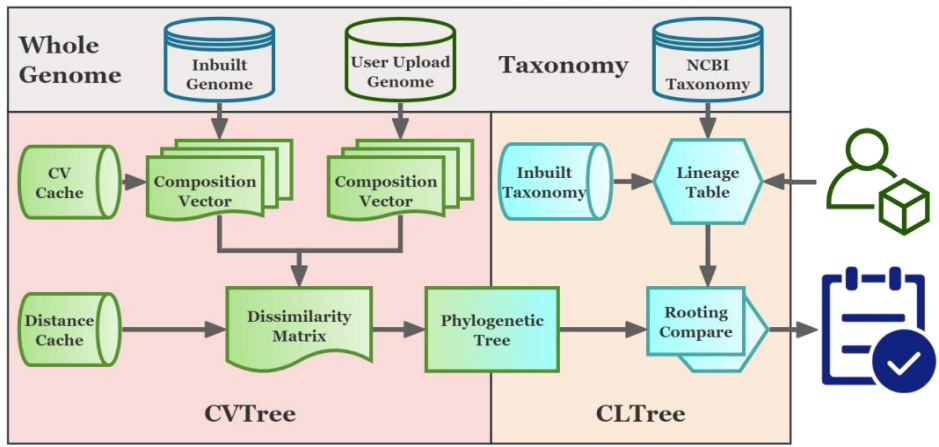

**Figure 1    Workflow of Backend of WebCVTree4**

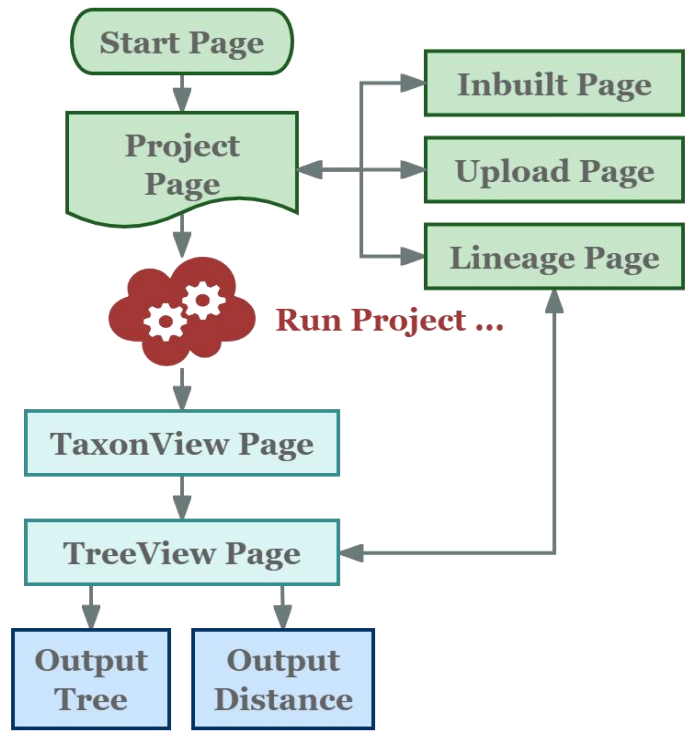

**Figure 2 Flow Diagram of Frontend of WebCVTree4**

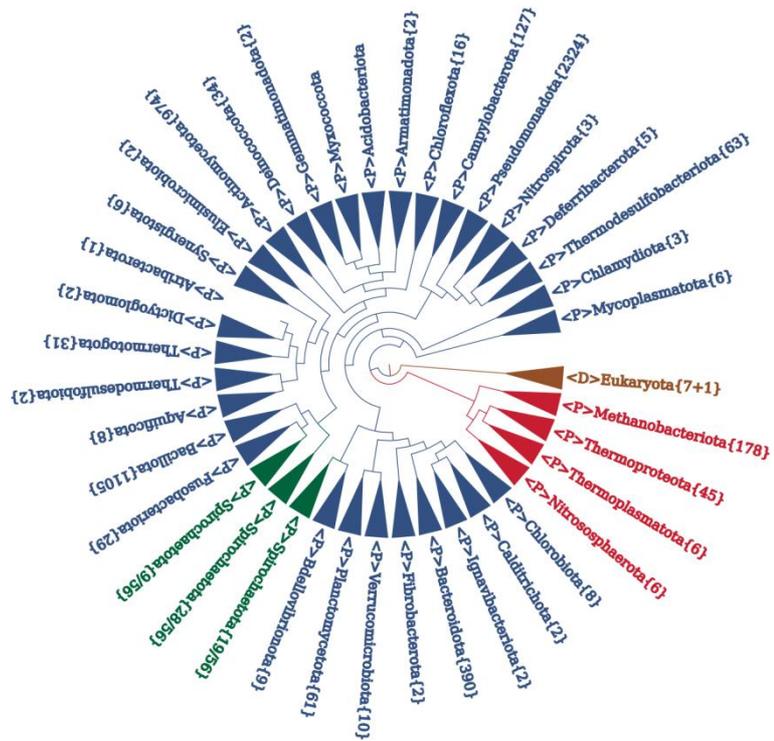

Figure 3 Example for Output Tree by WebCVTree4